\documentclass[preprint, showpacs,preprintnumbers,amsmath,amssymb,nofootinbib]{revtex4}

 \usepackage{epsf}
 \usepackage{graphicx}
\usepackage{float}
\begin{document}
\newcommand{\be}[1]{\begin{equation}\label{#1}}
 \newcommand{\ee}{\end{equation}}
 \newcommand{\bea}{\begin{eqnarray}}
 \newcommand{\eea}{\end{eqnarray}}
 \def\disp{\displaystyle}

 \def\gsim{ \lower .75ex \hbox{$\sim$} \llap{\raise .27ex \hbox{$>$}} }
 \def\lsim{ \lower .75ex \hbox{$\sim$} \llap{\raise .27ex \hbox{$<$}} }

\title{\Large \bf Examining the cosmic acceleration with the latest Union2 supernova data}
\author{Zhengxiang Li$^{1}$, Puxun Wu$^{2}$  and Hongwei Yu$^{1, 2} \footnote{Corresponding author: hwyu@hunnu.edu.cn}$ }

\address{$^1$Department of Physics and Key Laboratory of Low Dimensional Quantum Structures and Quantum
Control of Ministry of Education, Hunan Normal University, Changsha,
Hunan 410081, China
\\$^2$Center for Nonlinear Science and Department of Physics, Ningbo
University,  Ningbo, Zhejiang 315211, China }

\begin{abstract}
In this Letter, by reconstructing the $Om$ diagnostic and the
deceleration parameter $q$ from the latest Union2 Type Ia supernova
sample with and without the systematic error along with the baryon
acoustic oscillation (BAO) and the cosmic microwave background
(CMB), we study the cosmic expanding history, using the
Chevallier-Polarski-Linder (CPL) parametrization. We obtain that
Union2+BAO favor an expansion with a decreasing of the acceleration
at $z<0.3$. However, once the CMB data is added in the analysis, the
cosmic acceleration is found to be still increasing, indicating a
tension between low redshift data and high redshift one. In order to
reduce this tension significantly, two different methods are
considered and thus two different subsamples of Union2 are selected.
We then find that two different subsamples+BAO+CMB give completely
different results on the cosmic expanding history when the
systematic error is ignored, with one suggesting a decreasing cosmic
acceleration, the other just the opposite, although both of them
alone with BAO support that the cosmic acceleration is slowing down.
However,  once the systematic error is considered, two different
subsamples of Union2 along with BAO and CMB  all favor an increasing
of the present cosmic acceleration. Therefore a clear-cut answer on
whether the cosmic acceleration is slowing down calls for more
consistent data and more reliable methods to analyze them.

\end{abstract}


 \maketitle
 \renewcommand{\baselinestretch}{1.5}

\section{INTRODUCTION}\label{sec1}
The fact that our Universe has entered a state of accelerating
expansion at redshifts less than $\sim0.5$ is well established by
various independent observational data, including the Type Ia
Supernova (SNIa)~\cite{SNIa}, the large scale structure~\cite{BAO,
LSS}, the cosmic microwave background (CMB) radiation~\cite{CMB},
and so on. In order to explain this observed phenomena, one usually
assumes that there exists,  in our Universe,  an exotic energy
component, named dark energy (DE), which has  negative pressure and
thus can generate a repulsive force. It dominates our Universe and
drives it to an accelerating expansion at recent times. Since the
equation of state (EOS) $w$ of dark energy embodies its properties,
one may adopt  a parametrized form of $w(z)$ with several free
parameters, such as  the Chevallier-Polarski-Linder (CPL)
parametrization~\cite{CPL},  to probe the cosmic expanding history
and the evolutionary behavior of dark energy from observations.

However, the results are different, sometimes even contradictory,
when different observational data are used~\cite{slowing,
Guimaraes2010, Gong2010}. For example, by investigating the
diagnostic $Om$~\cite{diagnostic}, which is defined as
\begin{equation}
Om(z)\equiv\frac{E^2(z)-1}{(1+z)^3-1},~~~~~~E(z)=H(z)/H_0,
\end{equation}
and the deceleration parameter $q$ from the Constitution
SNIa~\cite{constitution} along with the baryonic acoustic
oscillation  (BAO) distance ratio data~\cite{bao1, bao2} and using
the CPL parametrization, Shafieloo et al.~\cite{slowing} found that
the cosmic expansion acceleration  might be slowing down, which is
different from studies with other SNIa data sets~\cite{Gong2010}.
However, once the CMB data is included, their result turns out to be
consistent with the $\Lambda$CDM model very well and the universe is
undergoing an accelerating expansion with an increasing
acceleration. So, there appears some tension between low redshift
data (Constitution SNIa+BAO) and high redshift (CMB) one.
Surprisingly, further analysis using a subsample (SNLS+ESSENCE+CfA)
of the Constitution SNIa reveals that the outcome that the cosmic
acceleration has been over the peak does not rely on whether the CMB
data is added,  and the tension between SNIa and CMB is reduced
significantly. Actually, although previous SNIa data sets, such as
Gold06~\cite{Riess2007} and Union~\cite{Kowalski2008}, do not
support the result that the cosmic acceleration is slowing down, the
tension between them and CMB has already been found
~\cite{Jassal2005, Nesseris2007}, and in Ref.~\cite{Nesseris2007},
Nesseris and Perivolaropoulos  proposed a simple  method  to find
the outliers responsible for it.

Recently, the largest and latest SNIa sample (Union2) was released
by the Supernova Cosmology Project (SCP)
Collaboration~\cite{Union2}. It consists of 557 data points. We list
the subsets in detail in Tab.~(\ref{Tab1}). In this Letter, we plan
to reexamine the cosmic expanding history from the Union2, BAO and
CMB by using the popular CPL parametrization. The tension between
low redshift data and high redshift one is also analyzed in detail.

\section{OBSERVATIONAL DATA}\label{sec2}
For SNIa data, we use the latest Union2 compilation released by the
Supernova Cosmology Project (SCP) Collaboration
recently~\cite{Union2}. It consists of 557 data points and is the
largest published SNIa sample up to now. The statistics of each
subset with $3\sigma$ outlier rejection are detailed in
Tab.~(\ref{Tab1}). We fit the SNIa with cosmological models by
minimizing the $\chi^2$ value of the distance modulus
\begin{equation}
\chi^2=\sum_{i,j=1}^{557}[\mu(z_i)-\mu_{obs}(z_i)]C_{sn}^{-1}(z_i,z_j)[\mu(z_j)-\mu_{obs}(z_j)],
\end{equation}
where  $\mu(z)\equiv 5\log_{10}[d_L(z)/Mpc]+25$ is the theoretical
value of the distance modulus, $\mu_{obs}$ is the corresponding
observed one, and $C_{sn}(z_i,z_j)$ is the covariance matrix, which
was detailed in Ref.~\cite{Union2} and  can be found on the web
site~\footnote{http://supernova.lbl.gov/Union/}. In the present
Letter, we will use two different covariance matrices, which
correspond to the cases with and without systematic error,
respectively.   The luminosity distance $d_L(z)$ is
\begin{equation}
d_L(z)=\frac{1+z}{H_0}\int_0^z\frac{dz'}{E(z')}
\end{equation}
For the CPL parametrization,  $w=w_0+w_1\frac{z}{1+z}$,
\begin{equation} E^2(z)=
\Omega_{0m}(1+z)^3+(1-\Omega_{0m})(1+z)^{3(1+w_0+w_1)}\exp\bigg(-\frac{3w_1z}{1+z}\bigg)\;,
\end{equation}
where $\Omega_{0m}$ is the present dimensionless density parameter
of matter.

Since $H_0$ is a nuisance parameter, we marginalize over it by
minimizing the following expression
\begin{equation}
\chi_{SNIa}^2=\sum_{i,j=1}^{557}\alpha_iC_{sn}^{-1}(z_i,z_j)\alpha_j-\frac{[\sum_{ij}\alpha_iC_{sn}^{-1}(z_i,z_j)-\ln10/5]^2}
 {\sum_{ij}C_{sn}^{-1}(z_i,z_j)}-2\ln\bigg(\frac{\ln10}{5}\sqrt{\frac{2\pi}{\sum_{ij}C_{sn}^{-1}(z_i,z_j)}}\bigg),
\end{equation}
to obtain the constraint from SNIa, where
$\alpha_i=\mu_{obs}-25-5\log_{10}[H_0d_L(z_i)]$.

The BAO data considered in our analysis is the distance ratio
obtained at $z=0.20$ and $z=0.35$ from the joint analysis of the 2dF
Galaxy Redsihft Survey and SDSS data~\cite{bao2}, which is a relatively model independent quantity
and can be expressed as
\begin{equation}
\frac{D_V(z=0.35)}{D_V(z=0.20)}=1.736\pm 0.065,
\end{equation}
with
\begin{equation}
D_V(z_{BAO})=\bigg[\frac{z_{BAO}}{H(z_{BAO})}\bigg(\int_0^{z_{BAO}}\frac{dz}{H(z)}\bigg)^2\bigg]^{1/3}.
\end{equation}
Performing $\chi^2$ statistics as follows
\begin{equation}
\chi_{BAO}^2=\frac{[D_V(z=0.35)/D_V(z=0.20)-1.736]^2}{0.065^2},
\end{equation}
one can obtain the constraint from BAO.
A result from the combination of SNIa and BAO is given by calculating $\chi_{SNIa}^2+\chi_{BAO}^2$.

Furthermore, in our analysis we add the CMB redshift parameter~\cite{AstO1},
which is the reduce distance at $z_{ls}=1090$
\begin{equation}
R=\sqrt{\Omega_{0m}}\int_0^{z_{ls}}\frac{dz}{E(z)}=1.71\pm 0.019.
\end{equation}
We also apply the $\chi^2$
\begin{equation}
\chi^2_{CMB}=\frac{[R-1.71]^2}{0.019^2}\;,
\end{equation}
to find out the result from CMB and the constraints from SNIa+BAO+CMB are given by $\chi_{SNIa}^2+\chi_{BAO}^2+\chi_{CMB}^2$.

\begin{table}[I]
\begin{tabular}{|c|c|c|c|}
\hline ~~~~&\multicolumn{3}{|c|}{$\sigma_{cut}=3$}\\ \cline{2-4}
{\bf Set}&~~{\bf N}~~~~~&~~${\bf \sigma_{sys}(68\%)}$~~~~~& ~${\bf RMS(68\%)}$~\\
\hline
Hamuy {\it et al.} (1996)~\cite{Hamuy} & 18 &$0.15_{-0.03}^{+0.05}$ & $0.17_{-0.03}^{+0.03}$\\
\hline
Krisciunas {\it et al.} (2005)~\cite{Krisciunas} & 6 &$0.04_{-0.04}^{+0.13}$ & $0.11_{-0.03}^{+0.03}$\\
\hline
Riess {\it et al.} (1999)~\cite{Riess1} & 11 &$0.15_{-0.03}^{+0.07}$ & $0.17_{-0.04}^{+0.03}$\\
\hline
Jha {\it et al.} (2006)~\cite{Jha2} & 15 &$0.21_{-0.04}^{+0.07}$ & $0.22_{-0.04}^{+0.04}$\\
\hline
Kowalski {\it et al.} (2008)~\cite{Kowalski} & 8 &$0.07_{-0.06}^{+0.09}$ & $0.15_{-0.04}^{+0.03}$\\
\hline
\underline{Hicken {\it et al.} (2009)~\cite{constitution}} & 102 &$0.15_{-0.01}^{+0.02}$ & $0.19_{-0.01}^{+0.01}$\\
\hline
\underline{Holtzman {\it et al.} (2009)~\cite{Holtzman}} & 129 &$0.10_{-0.01}^{+0.01}$ & $0.15_{-0.01}^{+0.01}$\\
\hline
~Riess {\it et al.} (1998) + HZT~\cite{SNIa}~ & 11 &$0.31_{-0.09}^{+0.19}$ & $0.52_{-0.12}^{+0.10}$\\
\hline
Perlmutter {\it et al.} (1999)~\cite{SNIa} & 33 &$0.41_{-0.09}^{+0.12}$ & $0.64_{-0.08}^{+0.07}$\\
\hline
Barris {\it et al.} (2004)~\cite{Barris}  & 19 &$0.18_{-0.10}^{+0.13}$ & $0.38_{-0.07}^{+0.06}$\\
\hline
\underline{Amanullah {\it et al.} (2008)~\cite{Amanullah}}  & 5 &$0.19_{-0.06}^{+0.21}$ & $0.21_{-0.07}^{+0.05}$\\
\hline
Knop {\it et al.} (2003)~\cite{Knop}  & 11 &$0.05_{-0.05}^{+0.10}$ & $0.15_{-0.02}^{+0.03}$\\
\hline
\underline{Astier {\it et al.} (2006)~\cite{Astier}}  & 72 &$0.13_{-0.02}^{+0.03}$ & $0.21_{-0.02}^{+0.02}$\\
\hline
\underline{Miknaitis {\it et al.} (2007)~\cite{Miknaitis}}  & 74 &$0.19_{-0.03}^{+0.04}$ & $0.29_{-0.02}^{+0.02}$\\
\hline
Tonry {\it et al.} (2003)~\cite{Tonry}  & 6 &$0.15_{-0.12}^{+0.21}$ & $0.23_{-0.07}^{+0.05}$\\
\hline
Riess {\it et al.} (2007)~\cite{Riess3}  & 31 &$0.16_{-0.05}^{+0.06}$ & $0.45_{-0.06}^{+0.05}$\\
\hline
\underline{Amanullah {\it et al.} (2010)~\cite{Union2}}  & 6 &$0.00_{-0.00}^{+0.00}$ & $0.00_{-0.00}^{+0.00}$\\
\hline
Total~~& \multicolumn{3}{|l|}{~~557}\\
\hline
\end{tabular}
\tabcolsep 0pt \caption{\label{Tab1} Statistics of each subset with
$3\sigma$ outlier rejection for Union2 compilation.  The Union2S
consists of the underlined subsets.  } \vspace*{5pt}
\end{table}

\section{RESULTS}\label{sec3}
We first investigate the constraints on model parameters and then
analyze the evolutionary behavior of the decelerating parameter and
$Om(z)$ to probe the properties of dark energy and the cosmic
expanding history.

\begin{center}
 \begin{figure}[htbp]
 \centering
\includegraphics[width=0.45\textwidth, height=0.45\textwidth]{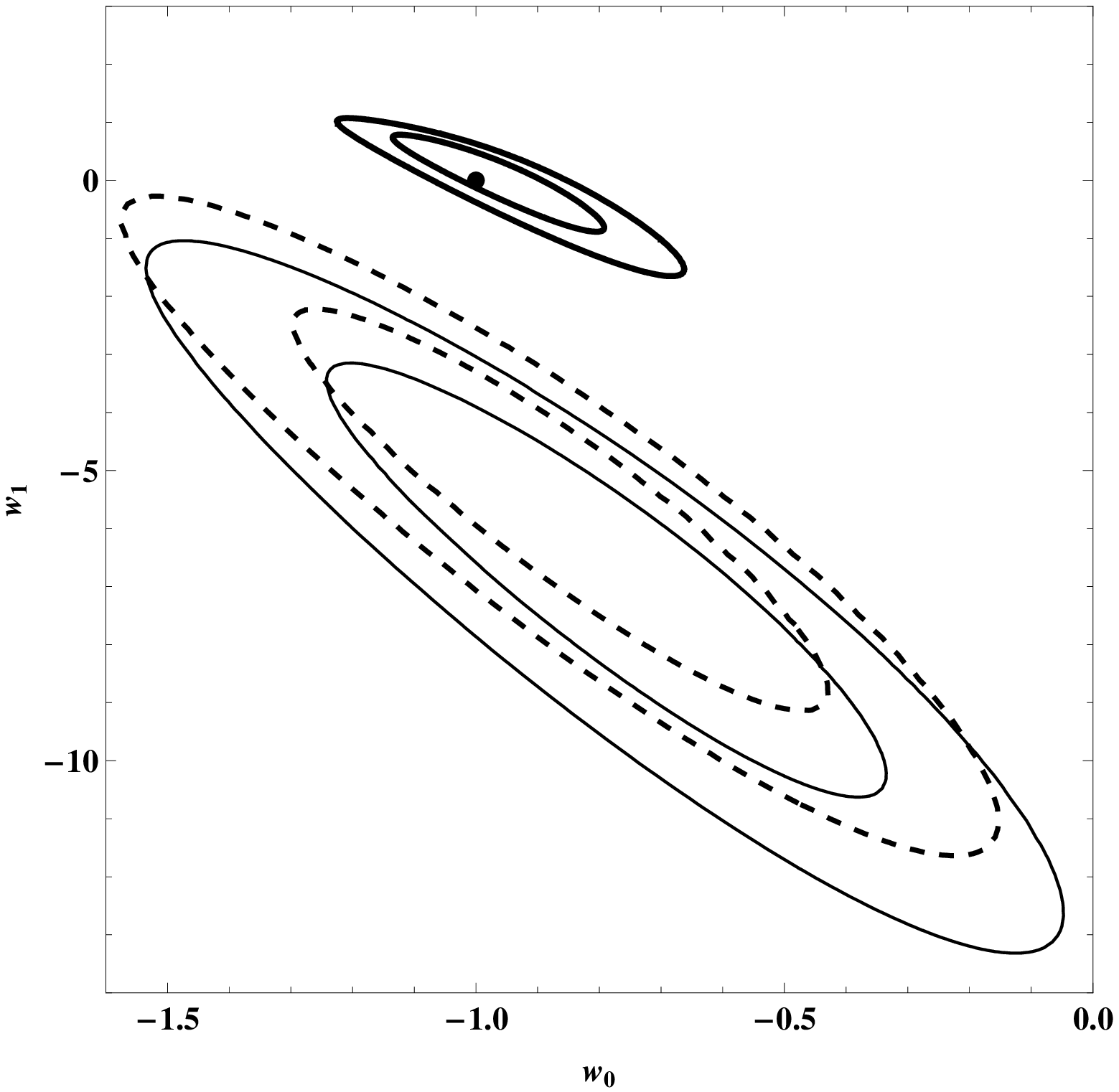}\includegraphics[width=0.45\textwidth, height=0.45\textwidth]{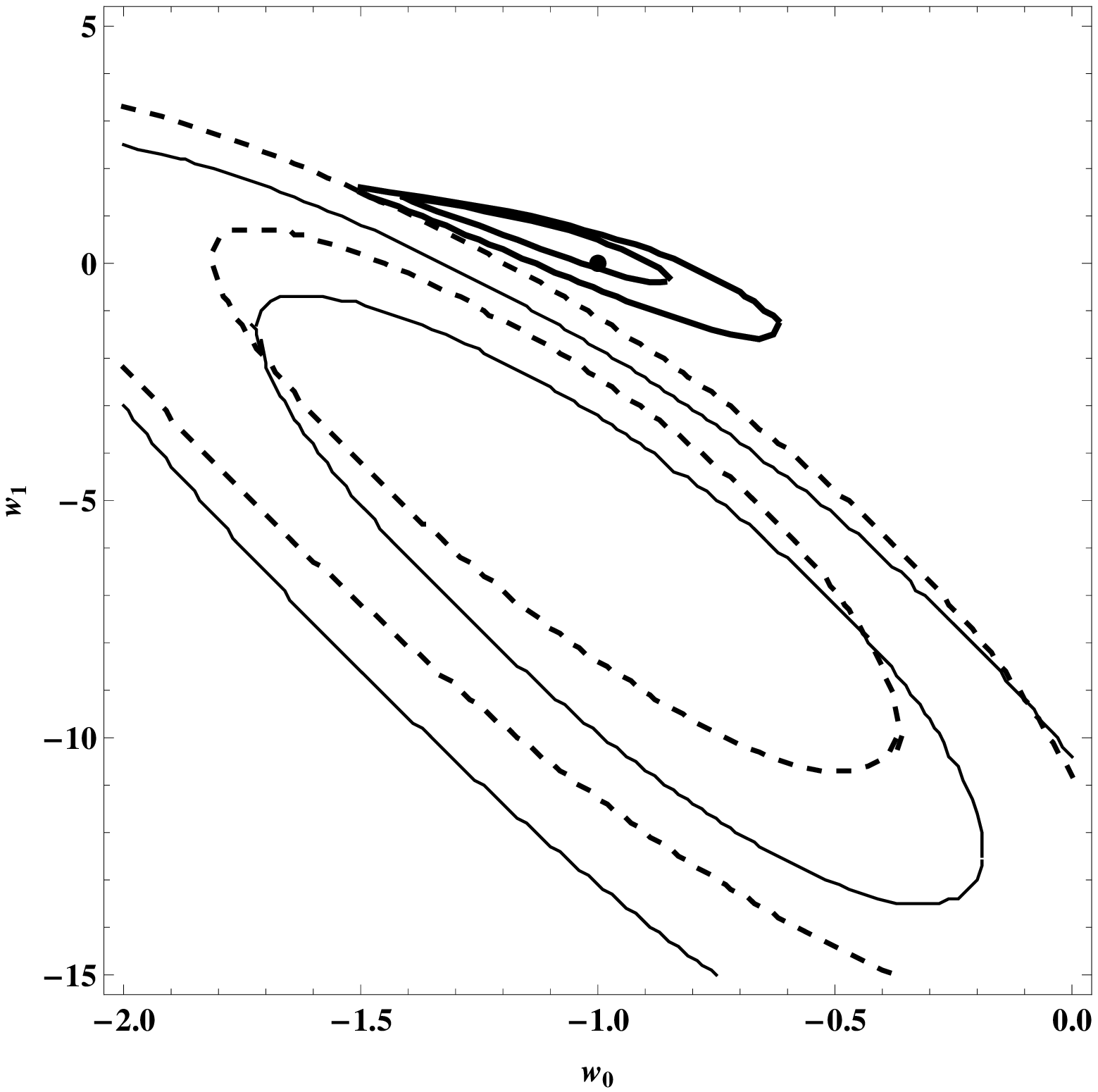}
 \caption{\label{Fig1}  The $68.3\%$ and $95\%$ confidence level regions
   for $w_0$ versus $w_1$. In  the left panel, the system error in the SNIa is ignored, while in the right panel,
   it is considered.  The dashed, solid   and
   thick solid lines represent the results obtained from
   Union2, Union2+BAO and Union2+BAO+CMB, respectively.
   The point at $w_0=-1$, $w_1=0$
   represents the spatially flat $\Lambda$CDM model.}
 \end{figure}
 \end{center}

Fig.~(\ref{Fig1}) shows the fitting contours of model parameters at
the $68.3\%$ and $95\%$ confidence levels. In the left panel, the
systematic error in the SNIa data is ignored, whereas in the right
panel, it is considered. The dashed, solid and thick solid lines
represent the results obtained from Union2, Union2+BAO and
Union2+BAO+CMB, respectively. The point at $w_0=-1$, $w_1=0$ denotes
the spatially flat $\Lambda$CDM model. We find that, independent of
whether the systematic  error is taken into account, the outcome
from Union2 is well consistent with that from Union2+BAO, and both
Union2 and Union2+BAO exclude the spatially flat $\Lambda$CDM
Universe at $95\%$ confidence level. However, compared to the good
overlap between regions  from Union2 and Union2+BAO, the one
obtained from Union2+BAO+CMB is relatively isolated and consistent
with the $\Lambda$CDM, which means that there exists  a tension
between low redshift data and high redshift one. Obviously, if we
use the SNIa with the systematic error, this tension is weaker than
that from the SNIa without.  That is, a consideration of systematic
errors in the SNIa alleviates this tension markedly.

\begin{center}
 \begin{figure}[htbp]
 \centering
\includegraphics[width=0.45\textwidth, height=0.45\textwidth]{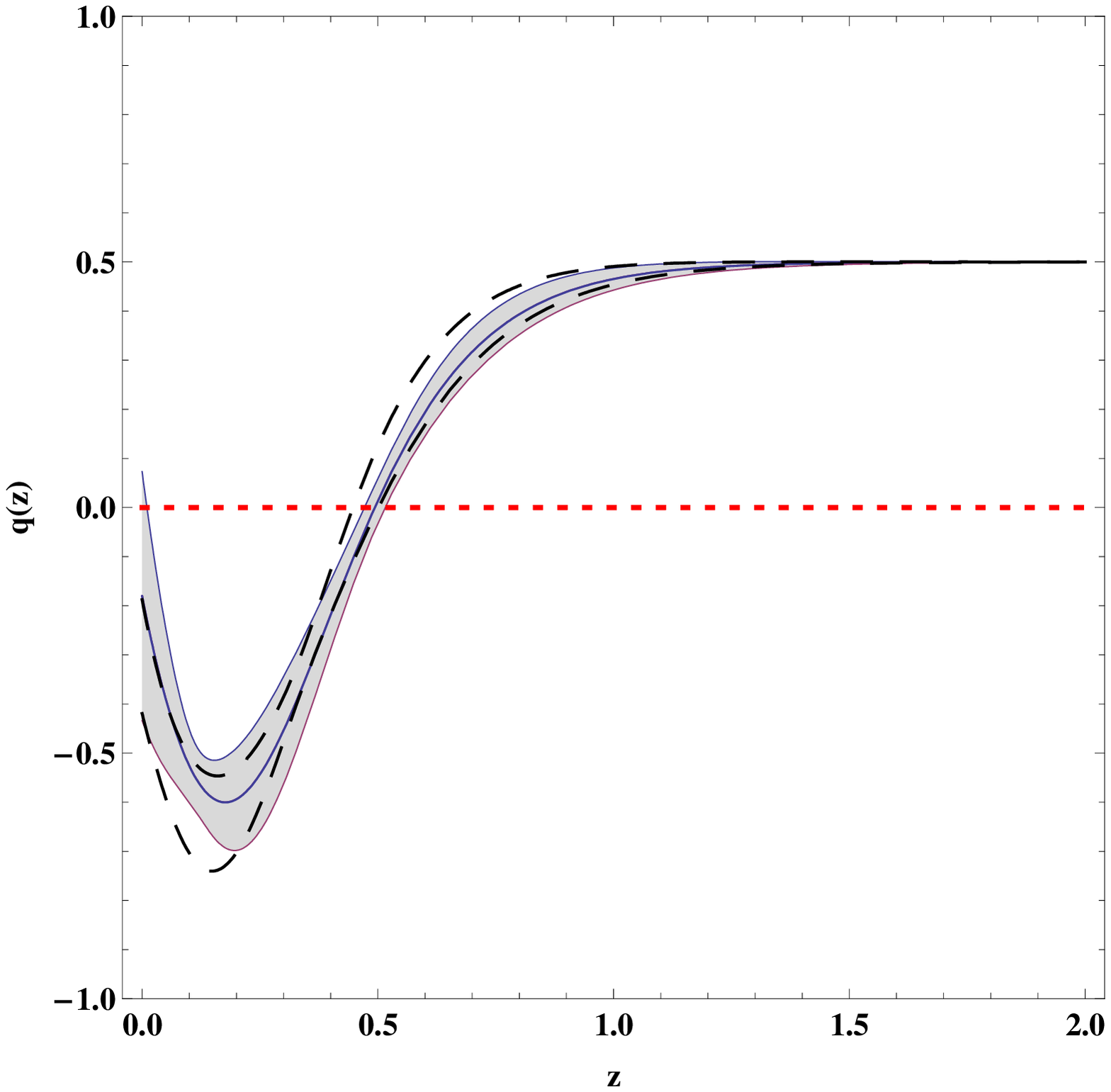}\includegraphics[width=0.45\textwidth, height=0.45\textwidth]{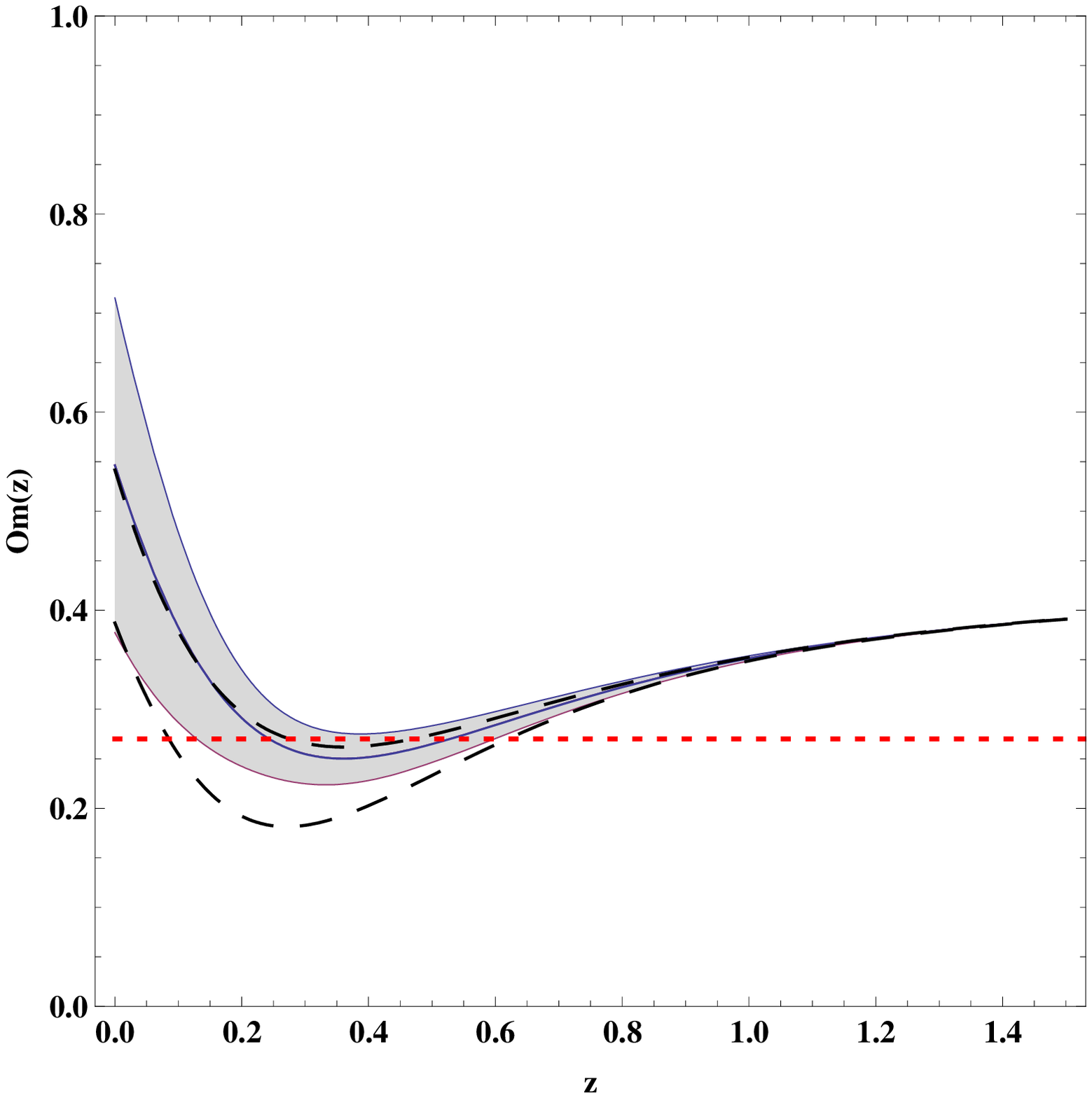}
\includegraphics[width=0.45\textwidth, height=0.45\textwidth]{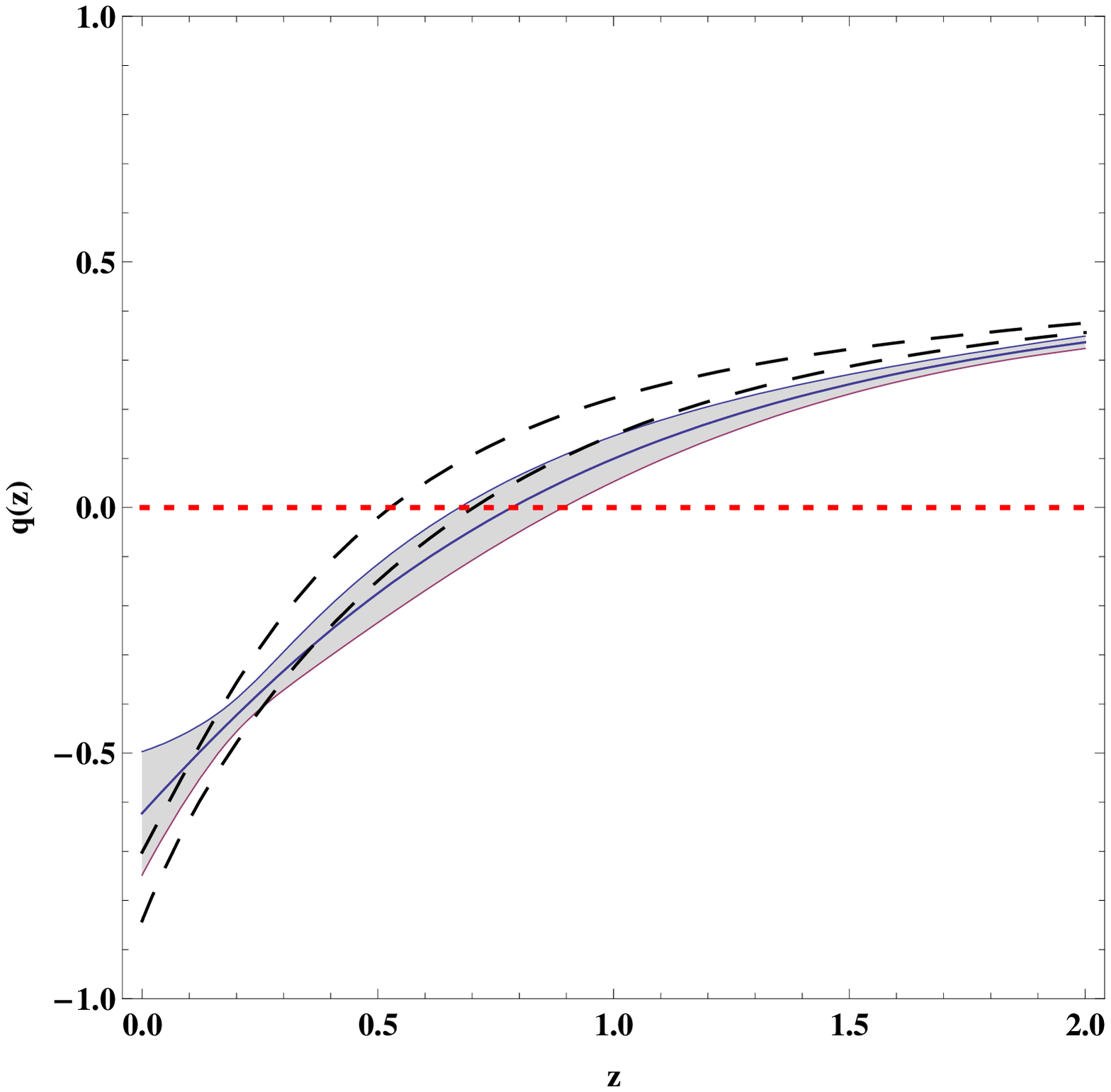}\includegraphics[width=0.45\textwidth, height=0.45\textwidth]{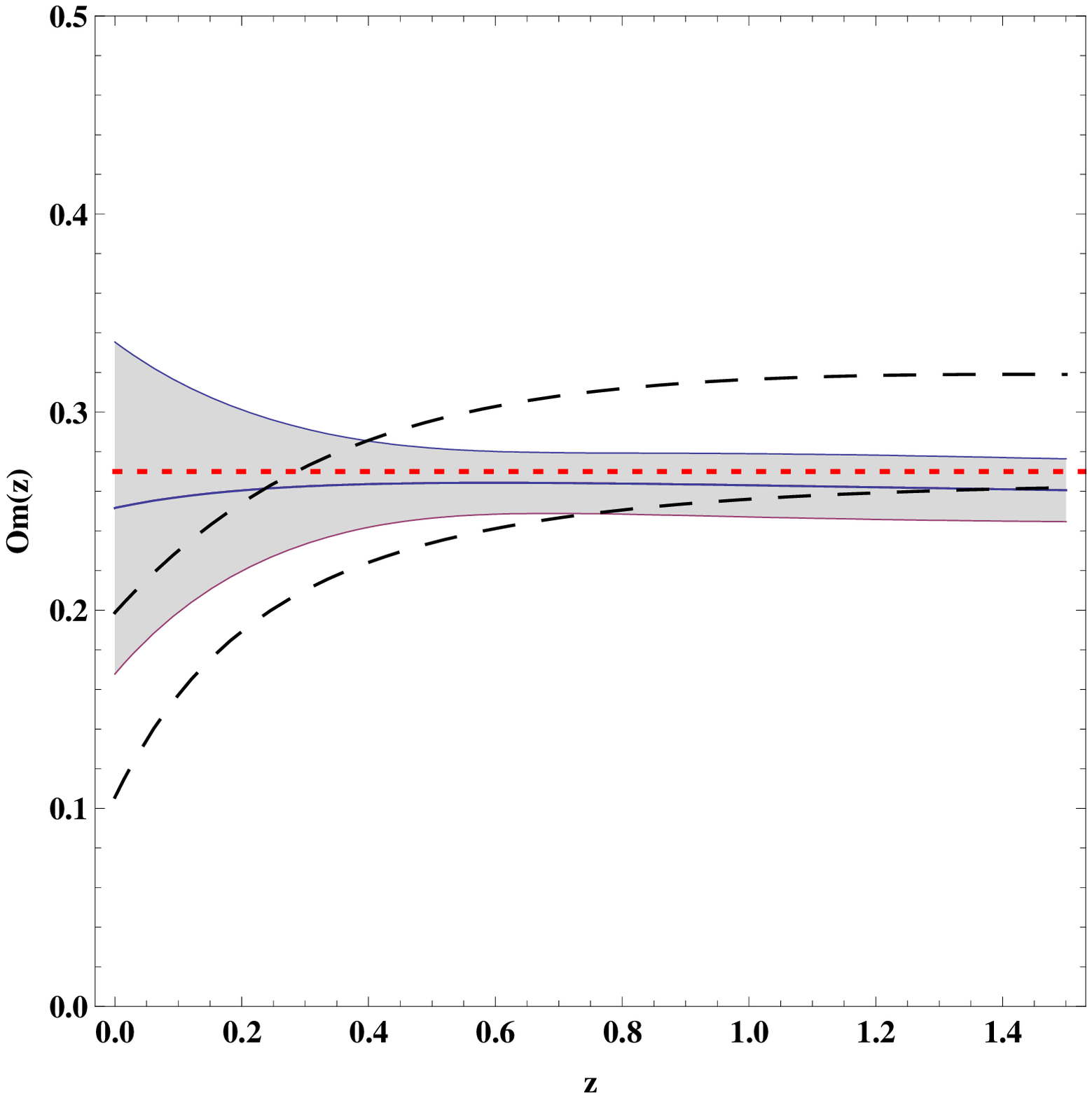}
 \caption{\label{Fig2}
The evolutionary behaviors of $q(z)$ and $Om(z)$ at the $68.3\%$
confidence level. The gray regions and the regions between two long
dashed lines  show the results without and with the systematic
errors in the SNIa, respectively.   The upper and lower panels
 represent the results reconstructed  from Union2+BAO and Union2+BAO+CMB, respectively.}
 \end{figure}
 \end{center}

The evolutionary behaviors of  $q(z)$ and $Om(z)$ at the $68.3\%$
confidence level reconstructed from Union2+BAO (upper panels) and
Union2+BAO+CMB (lower panels) are shown in Fig.~(\ref{Fig2}). The
gray regions and the regions between two long dashed lines represent
the results without and with the systematic errors in the SNIa,
respectively. It is easy to see that, for both the SNIa with
systematic error and without, there is an apparent rise of the
values of $Om(z)$ and $q(z)$ in redshifts $z<0.3$ for Union2+BAO
(upper panels), which means that the cosmic acceleration is slowing
down. However, this result changes dramatically with the addition of
CMB in the analysis, as shown in the lower panels of
Fig.~(\ref{Fig2}), which still supports an expansion with an
increasing acceleration. These results  are the same as that derived
from the Constitution SN Ia~\cite{slowing}.

\begin{center}
 \begin{figure}[htbp]
 \centering
\includegraphics[width=0.45\textwidth, height=0.45\textwidth]{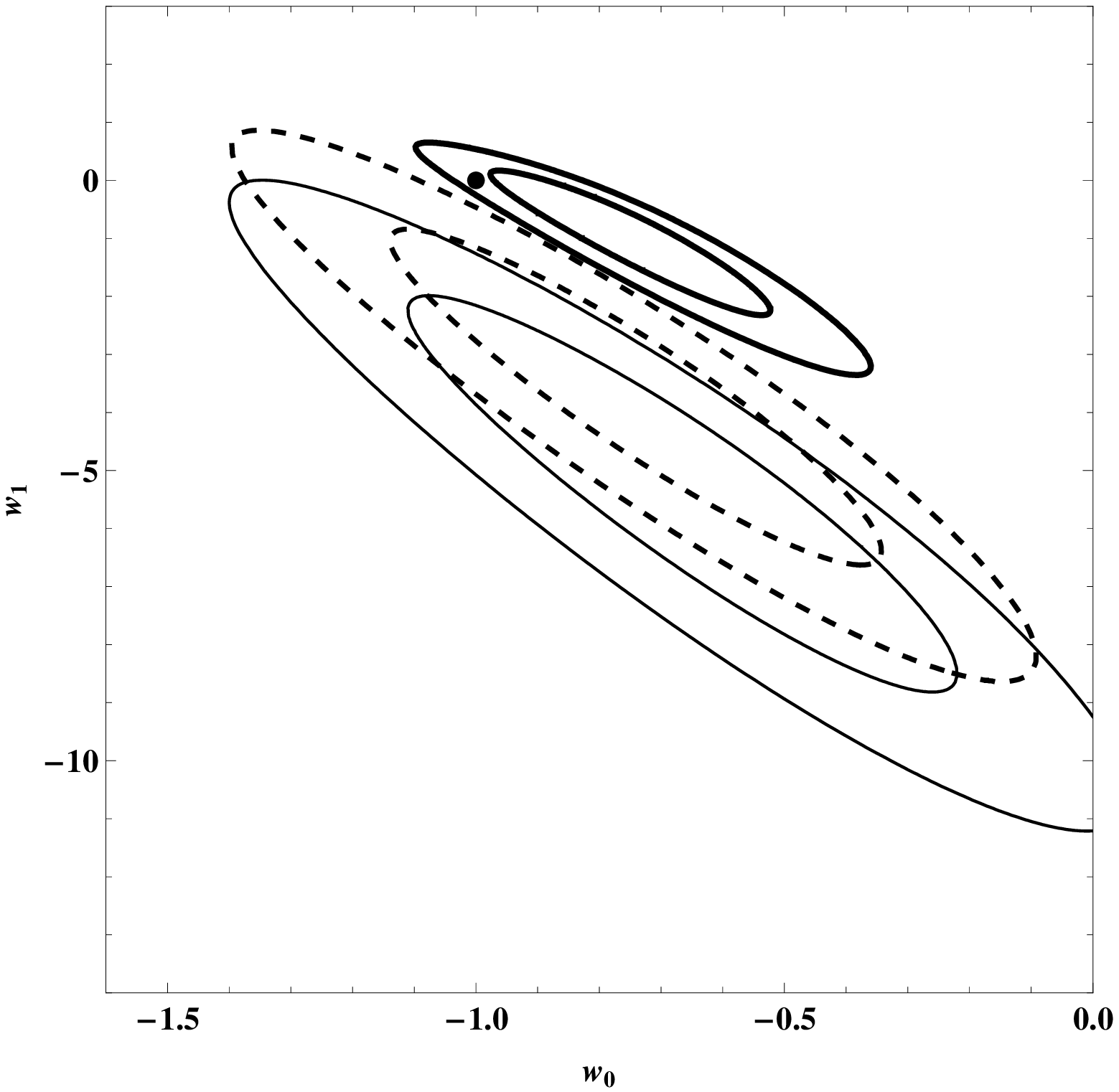}\includegraphics[width=0.45\textwidth, height=0.45\textwidth]{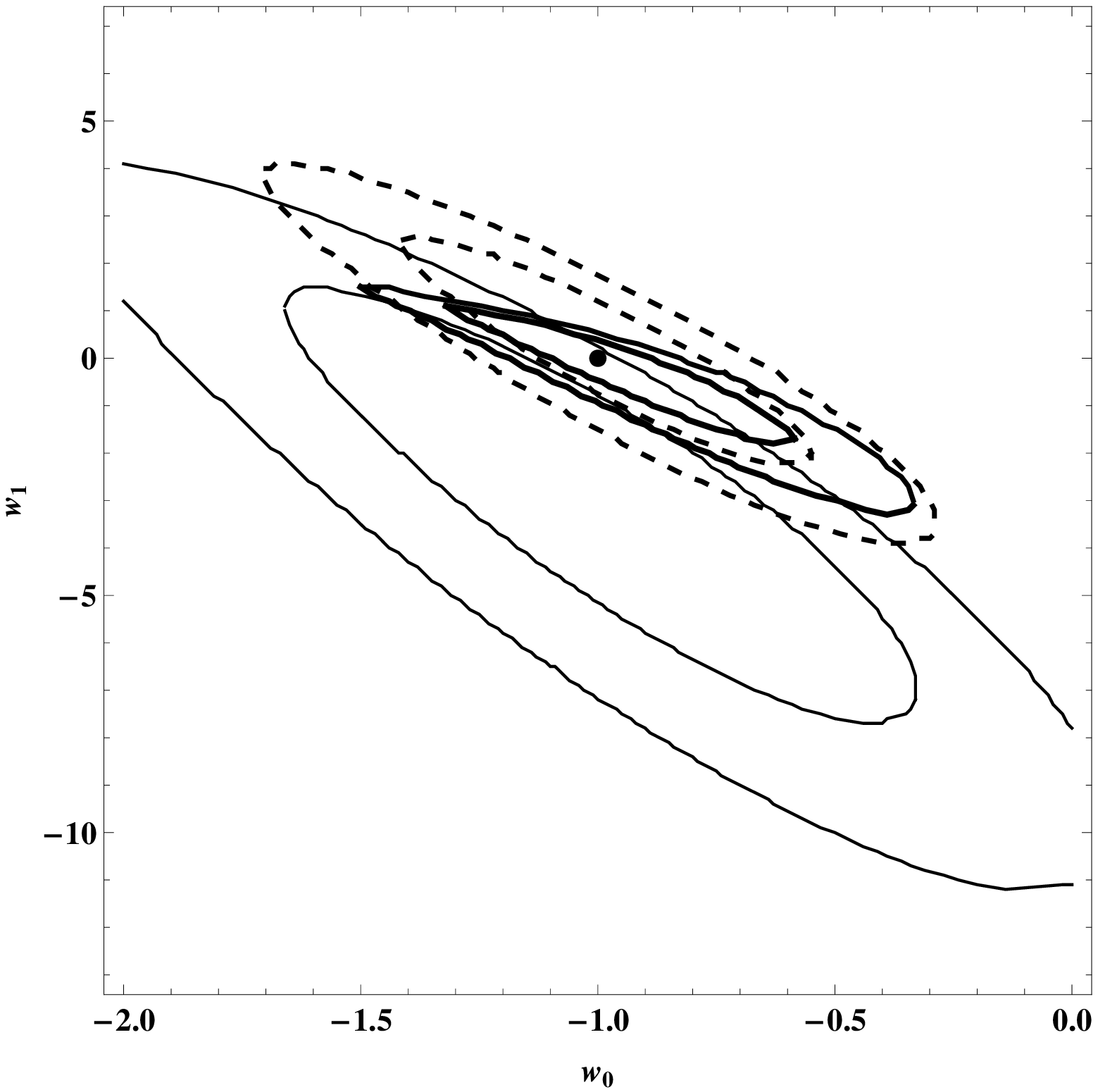}
 \caption{\label{Fig3}  The $68.3\%$ and $95\%$ confidence level regions
for $w_0$ versus $w_1$. A subsample (Union2S) of Union2 obtained
with the method in \cite{slowing} is considered. In  the left panel,
the system error in the SNIa is ignored, while in the right panel,
   it is considered.  The dashed, solid   and
   thick solid lines represent the results obtained from
   Union2S, Union2S+BAO and Union2S+BAO+CMB, respectively.
   The point at $w_0=-1$, $w_1=0$
   represents the spatially flat $\Lambda$CDM model.  }
 \end{figure}
 \end{center}

\begin{center}
\begin{figure}[htbp]
 \centering
\includegraphics[width=0.45\textwidth, height=0.45\textwidth]{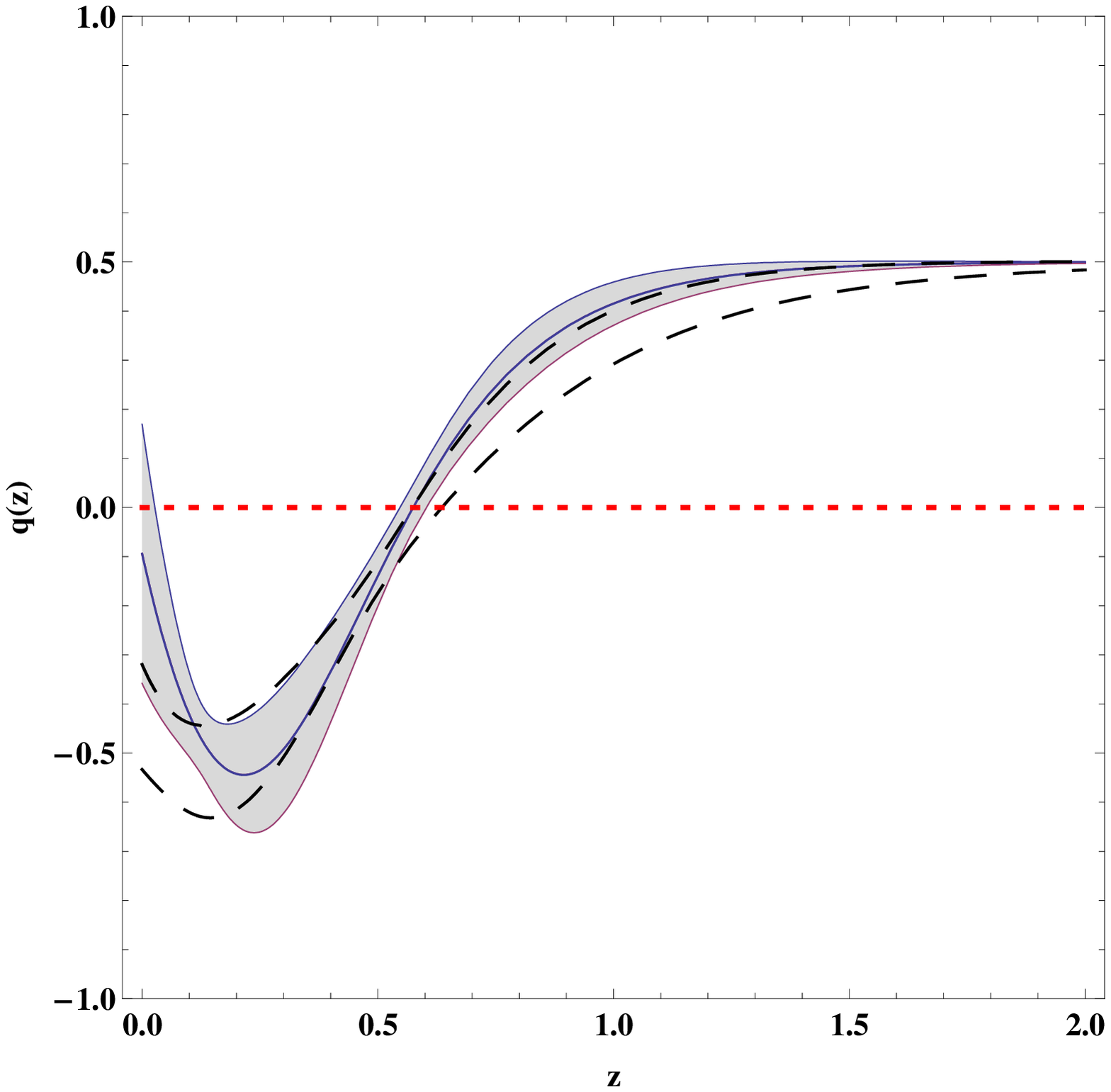}\includegraphics[width=0.45\textwidth, height=0.45\textwidth]{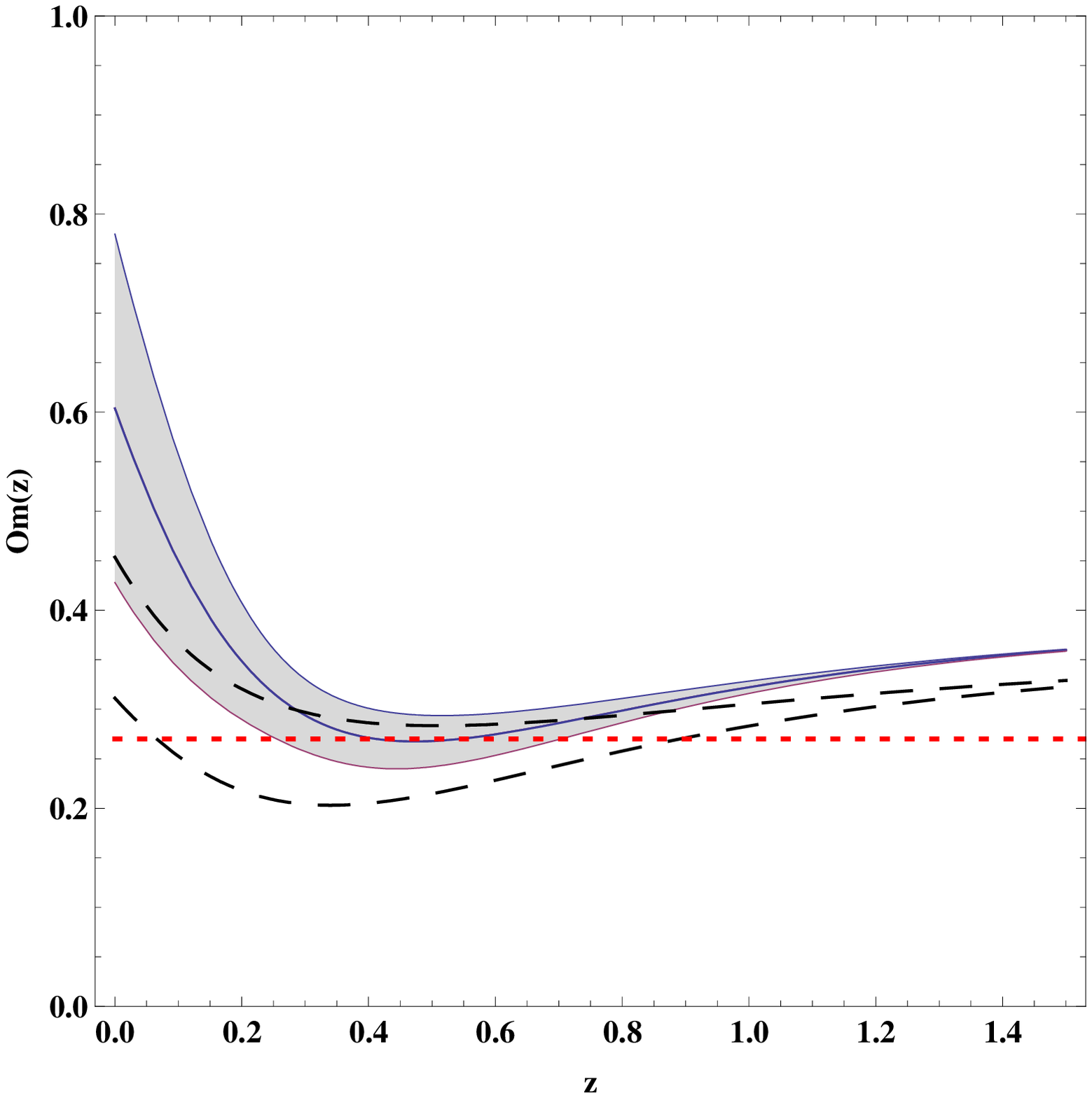}
\includegraphics[width=0.45\textwidth, height=0.45\textwidth]{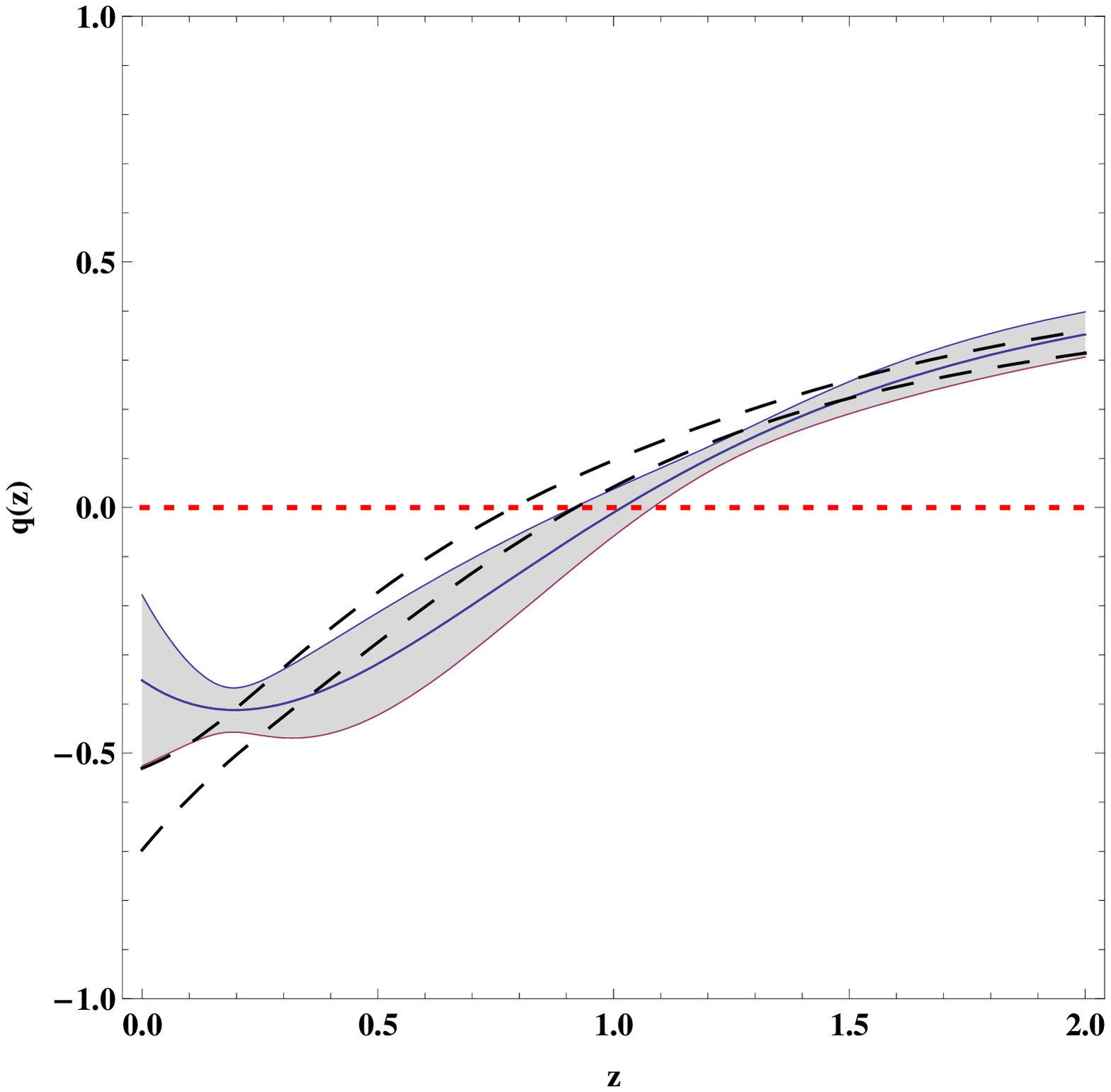}\includegraphics[width=0.45\textwidth, height=0.45\textwidth]{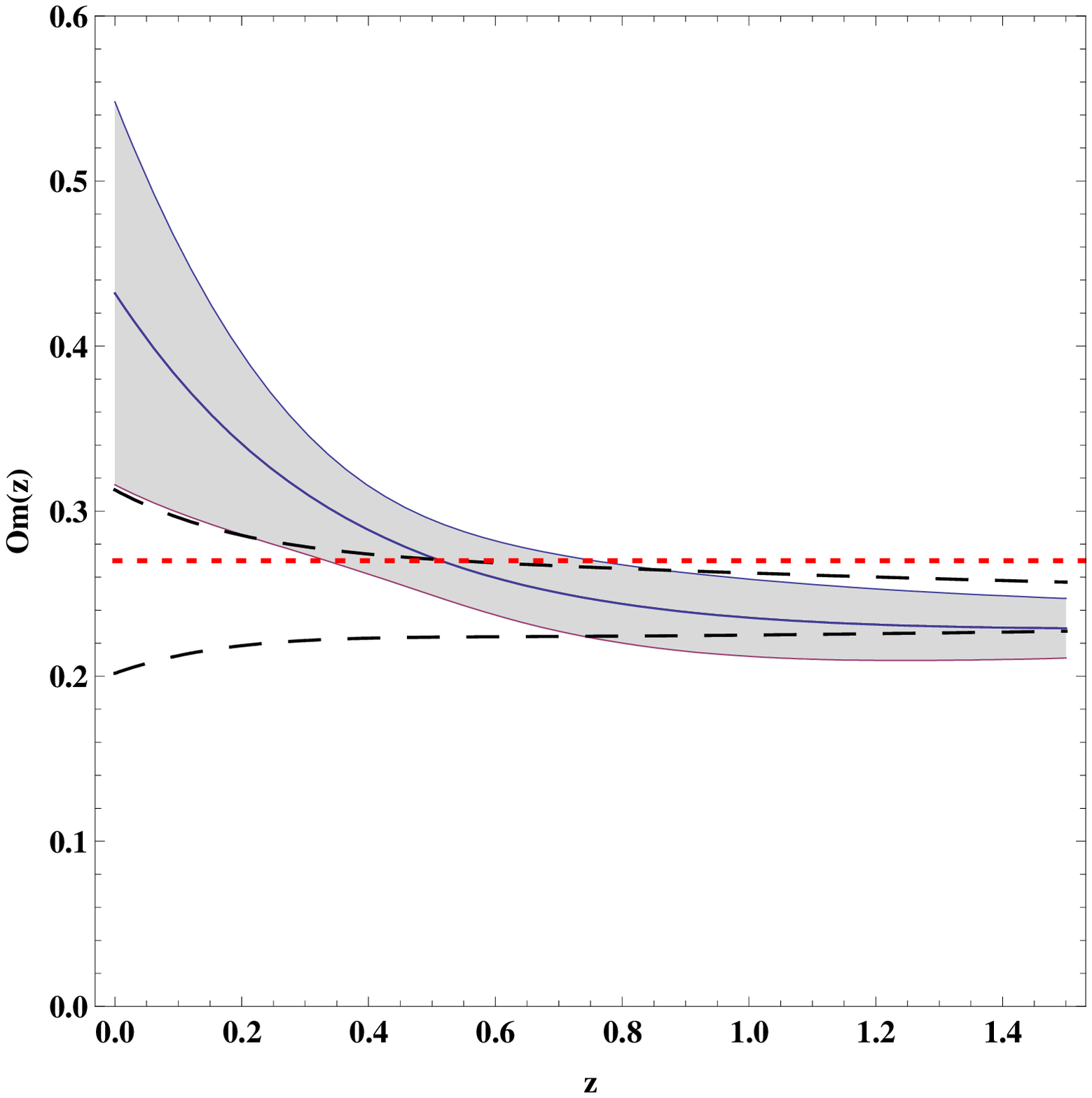}
 \caption{\label{Fig4}
 The evolutionary behaviors of $q(z)$ and $Om(z)$ at the $68.3\%$
confidence level. A subsample (Union2S) of Union2 obtained with the
method in \cite{slowing} is considered.  The gray regions and the
regions between two long dashed lines  show the results without and
with the systematic errors in the SNIa, respectively.   The upper
and lower panels
 represent the results reconstructed  from Union2S+BAO and Union2S+BAO+CMB, respectively. }
 \end{figure}
 \end{center}

In order to reduce the tension between low redshift data and high
redshift one, Shafieloo et al.~\cite{slowing} use a subsample of
Constitution SNIa sample, which is obtained by excluding the Gold
data, the high $z$ Hubble Space Telescope data and older SNIa data
sets in Constitution and thus it consists only of SNLS, ESSENCE and
CfA. They found that the tension is reduced significantly,  the
outcome does not rely on whether the CMB data is added and the
cosmic acceleration has been over the peak.  Here, we do a similar
analysis as that in Ref.~\cite{slowing} by using a subsample of the
Union2. This subsample, labeled as ``Union2S", could be obtained by
excluding the Gold data, the high $z$ Hubble Space Telescope data
and older SNIa data sets in the Union2. It contains   388 data
points and is given in detail in Tab.~(\ref{Tab1}) (underlined
subsets). The fitting contours for $w_0-w_1$ and reconstructed
$q(z)$ and $Om(z)$ are shown in Figs.~(\ref{Fig3},\ref{Fig4}),
respectively. From Fig.~(\ref{Fig3}), one can see that the tension
between low redshift data and high redshift one is reduced
noticeably, and the $\Lambda$CDM is consistent with Union2S with
systematic error (Union2S(sys)) and Union2S(sys)+BAO+CMB at the
$68\%$ confidence level. The left panel of Fig.~(\ref{Fig4}) shows
that, for the case with the systematic error in the SNIa ignored,
the evolution of $q(z)$ and $Om(z)$ reconstructed using Union2S+BAO
is similar to that from Union2S+BAO+CMB. Both of them favor that the
cosmic acceleration is slowing down. So, the same conclusion as that
 from the Constitution SNIa~\cite{slowing} is obtained.
However, once the systematic error in the SNIa is considered, the
results from Union2S(sys)+BAO+CMB show that the peak of $q(z)$ at
$z<0.3$  disappears, although Union2S(sys)+BAO still favor a slowing
down of the present cosmic acceleration.

\begin{center}
 \begin{figure}[htbp]
 \centering
\includegraphics[width=0.45\textwidth, height=0.45\textwidth]{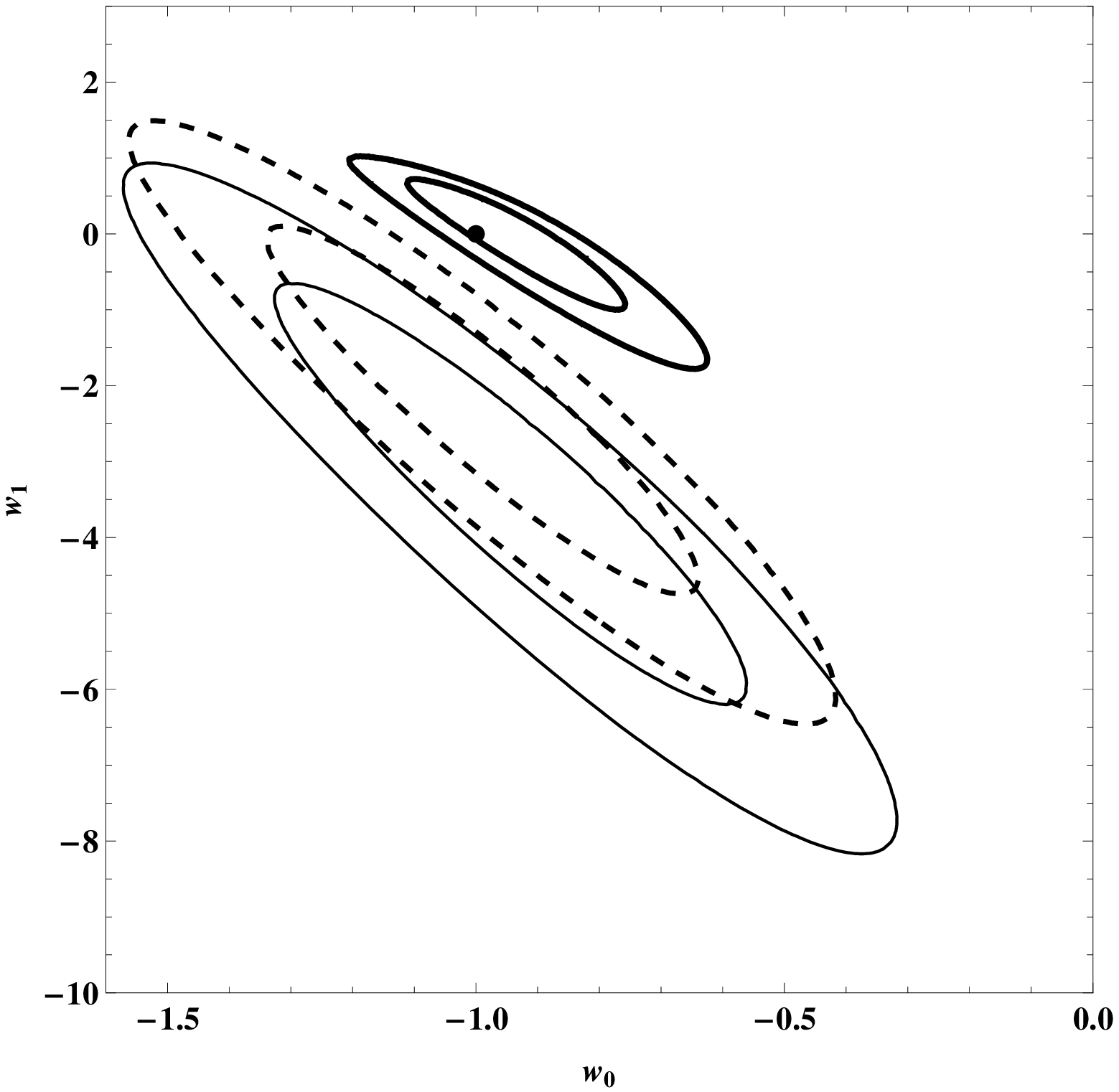}\includegraphics[width=0.45\textwidth, height=0.45\textwidth]{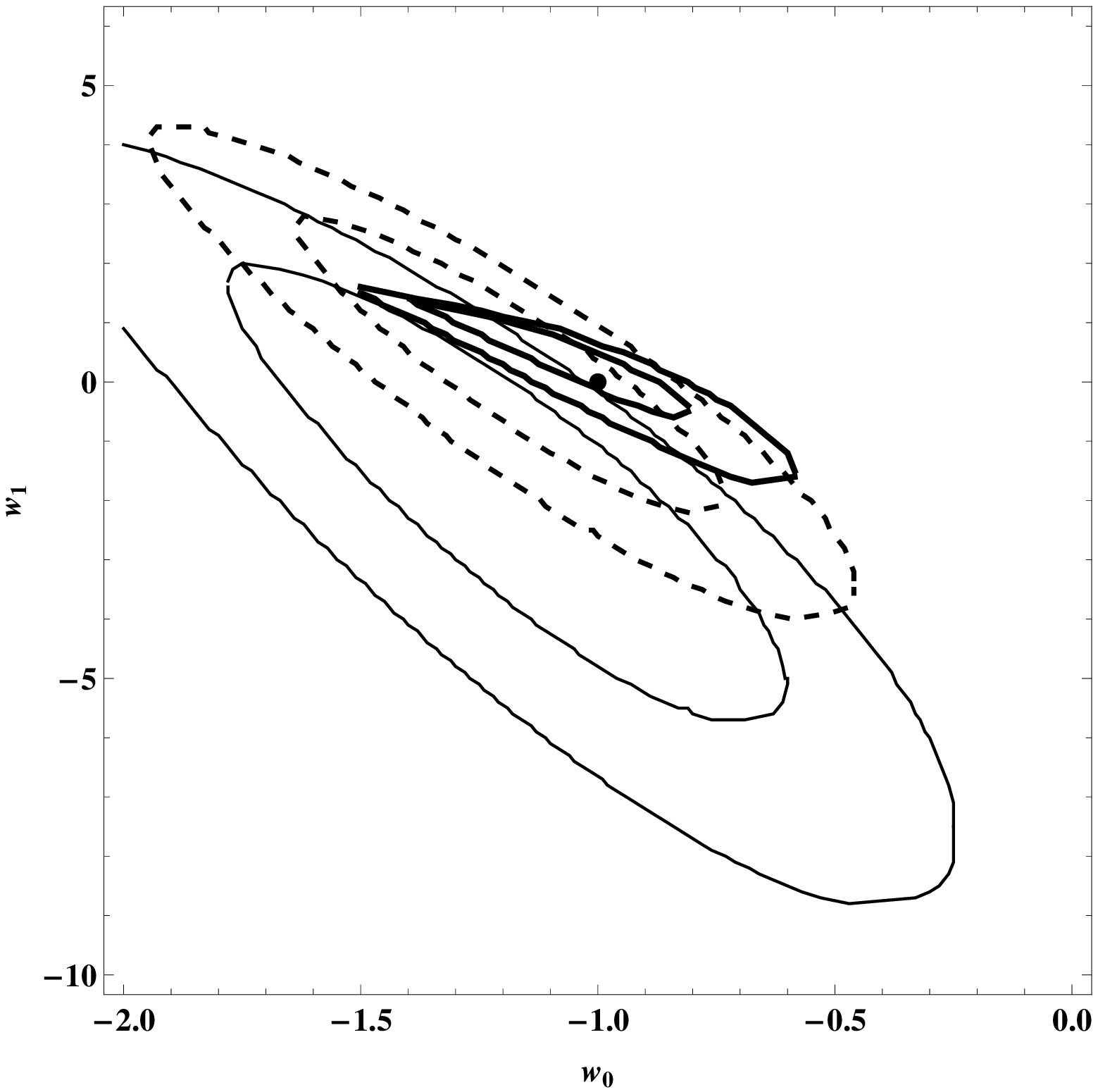}
 \caption{\label{Fig5}  The $68.3\%$ and $95\%$ confidence level regions
   for $w_0$ versus $w_1$. A subsample (Union2T) of Union2 obtained with the method in \cite{Nesseris2007} is considered.
 In  the left panel, the system error in the SNIa is ignored, while in the right panel,
   it is considered.  The dashed, solid   and
   thick solid lines represent the results obtained from
   Union2T, Union2T+BAO and Union2T+BAO+CMB, respectively.
   The point at $w_0=-1$, $w_1=0$
   represents the spatially flat $\Lambda$CDM model.  }
 \end{figure}
 \end{center}

\begin{center}
\begin{figure}[htbp]
 \centering
\includegraphics[width=0.45\textwidth, height=0.45\textwidth]{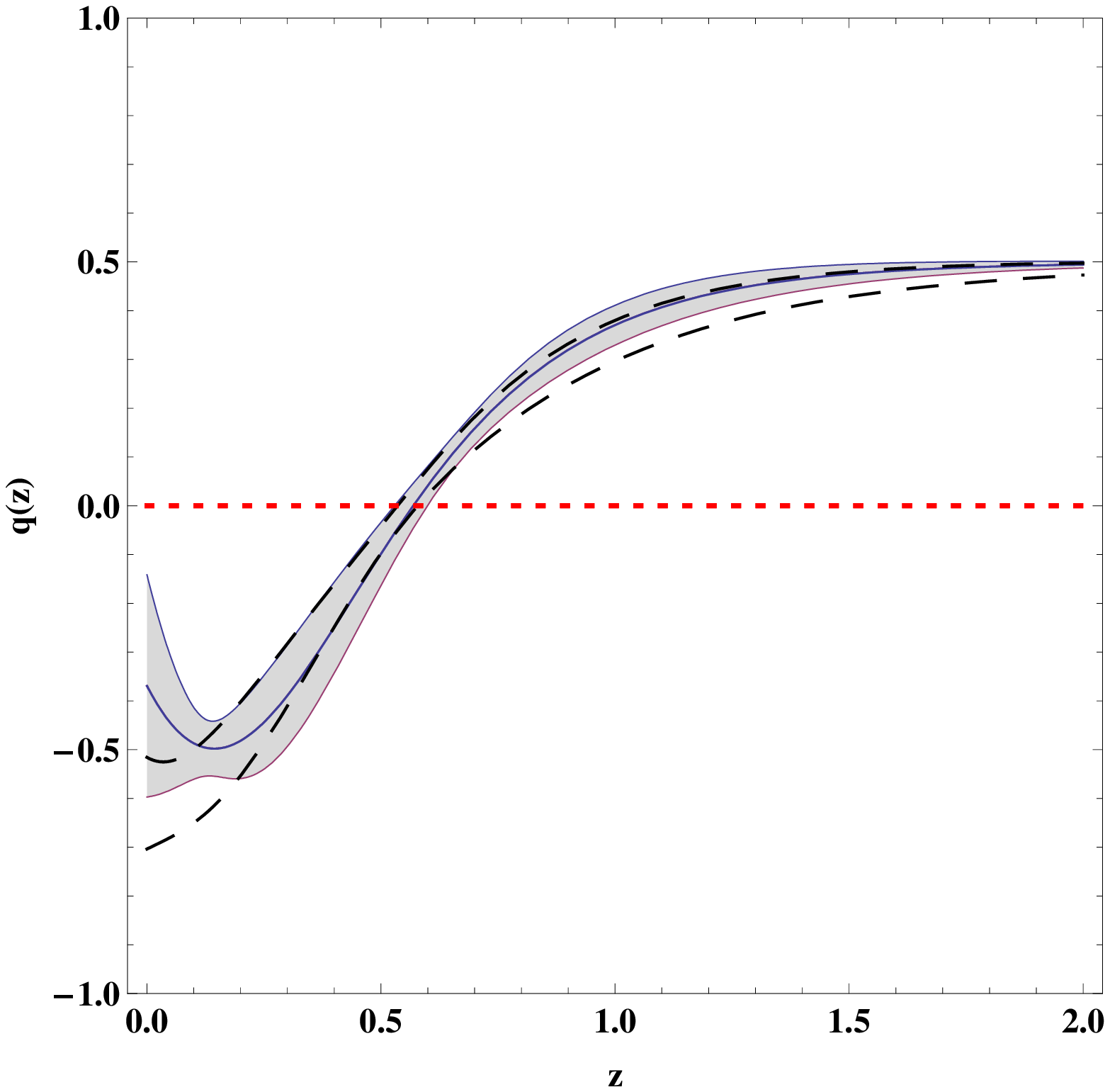}\includegraphics[width=0.45\textwidth, height=0.45\textwidth]{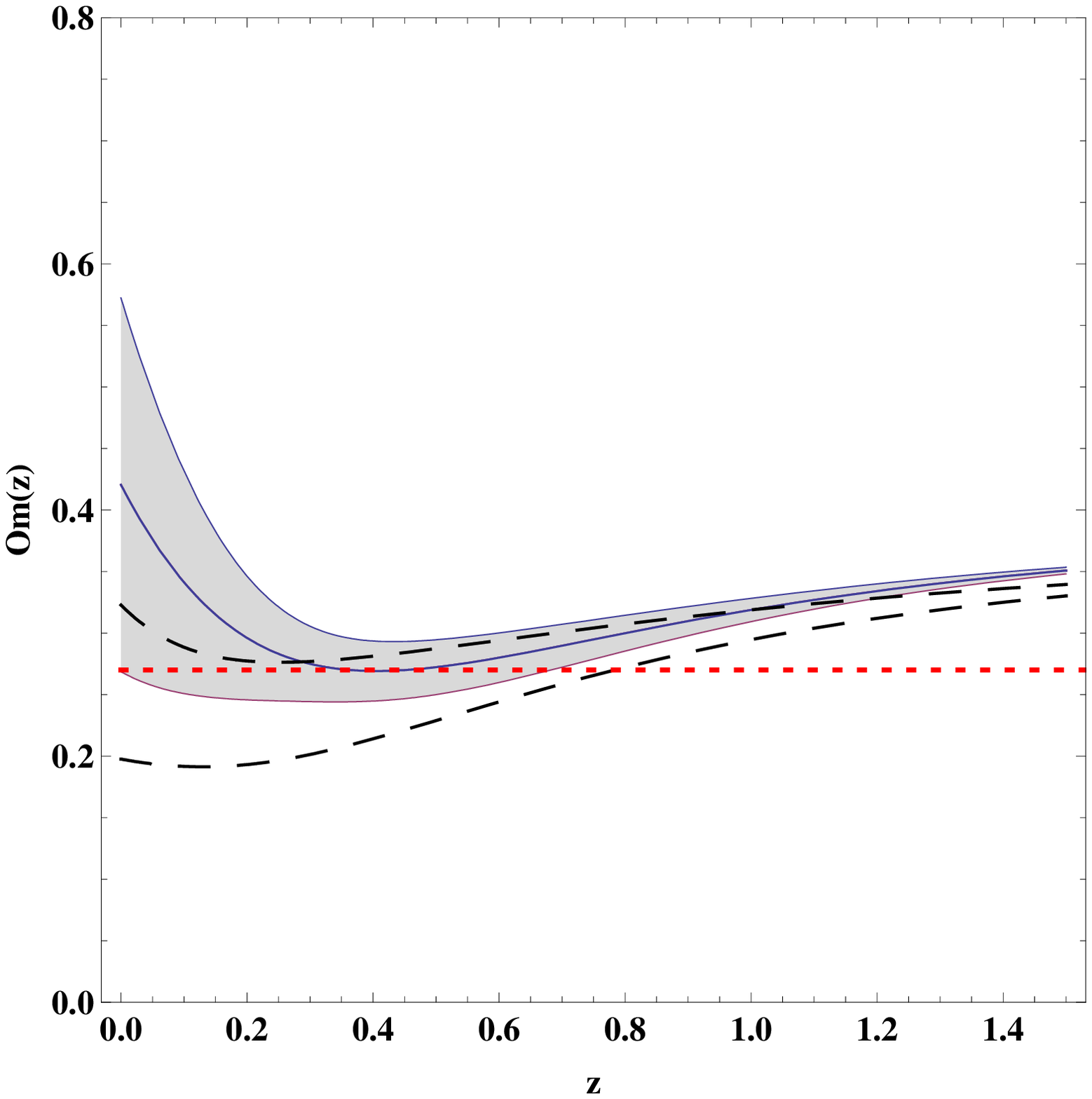}
\includegraphics[width=0.45\textwidth, height=0.45\textwidth]{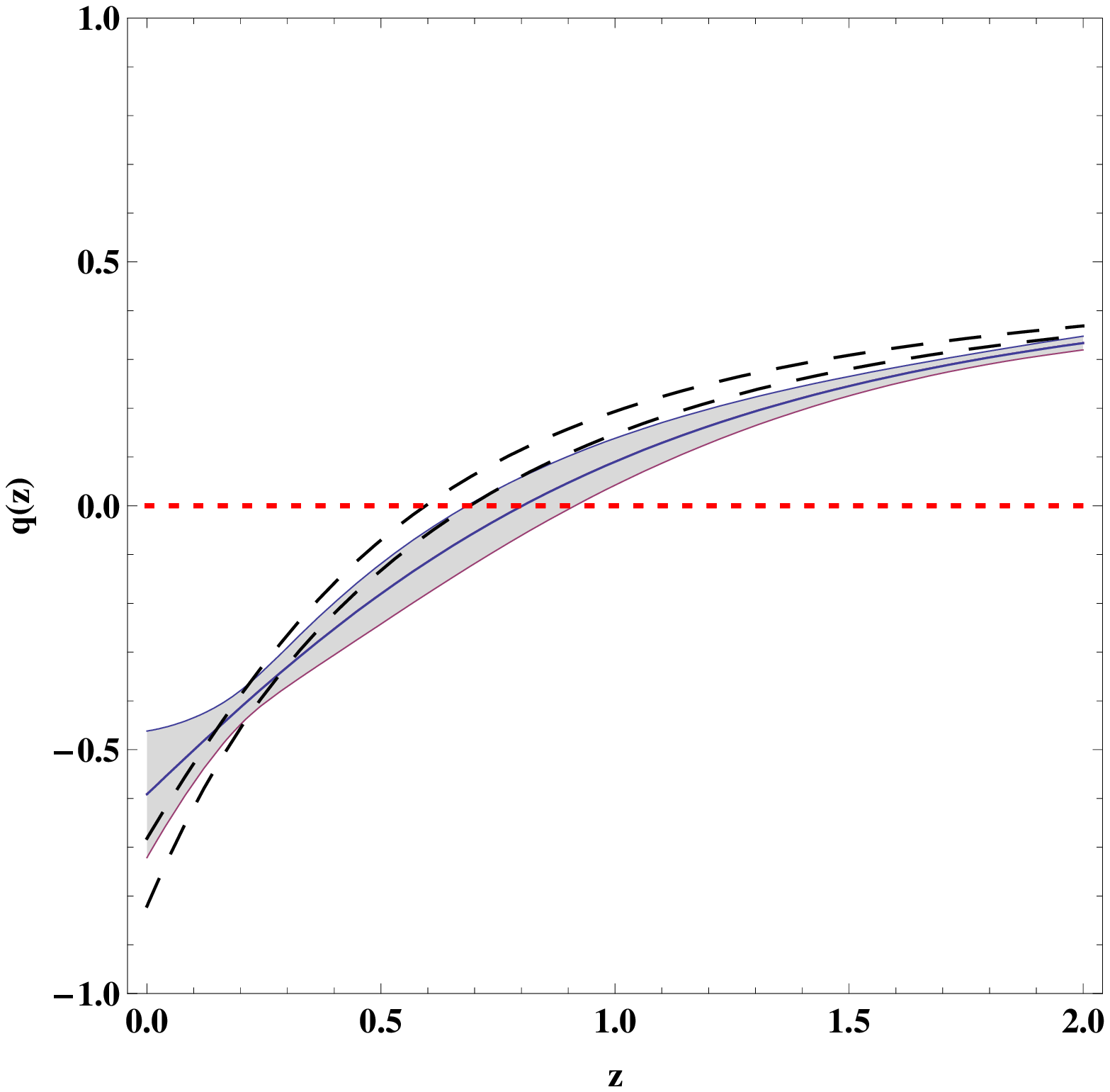}\includegraphics[width=0.45\textwidth, height=0.45\textwidth]{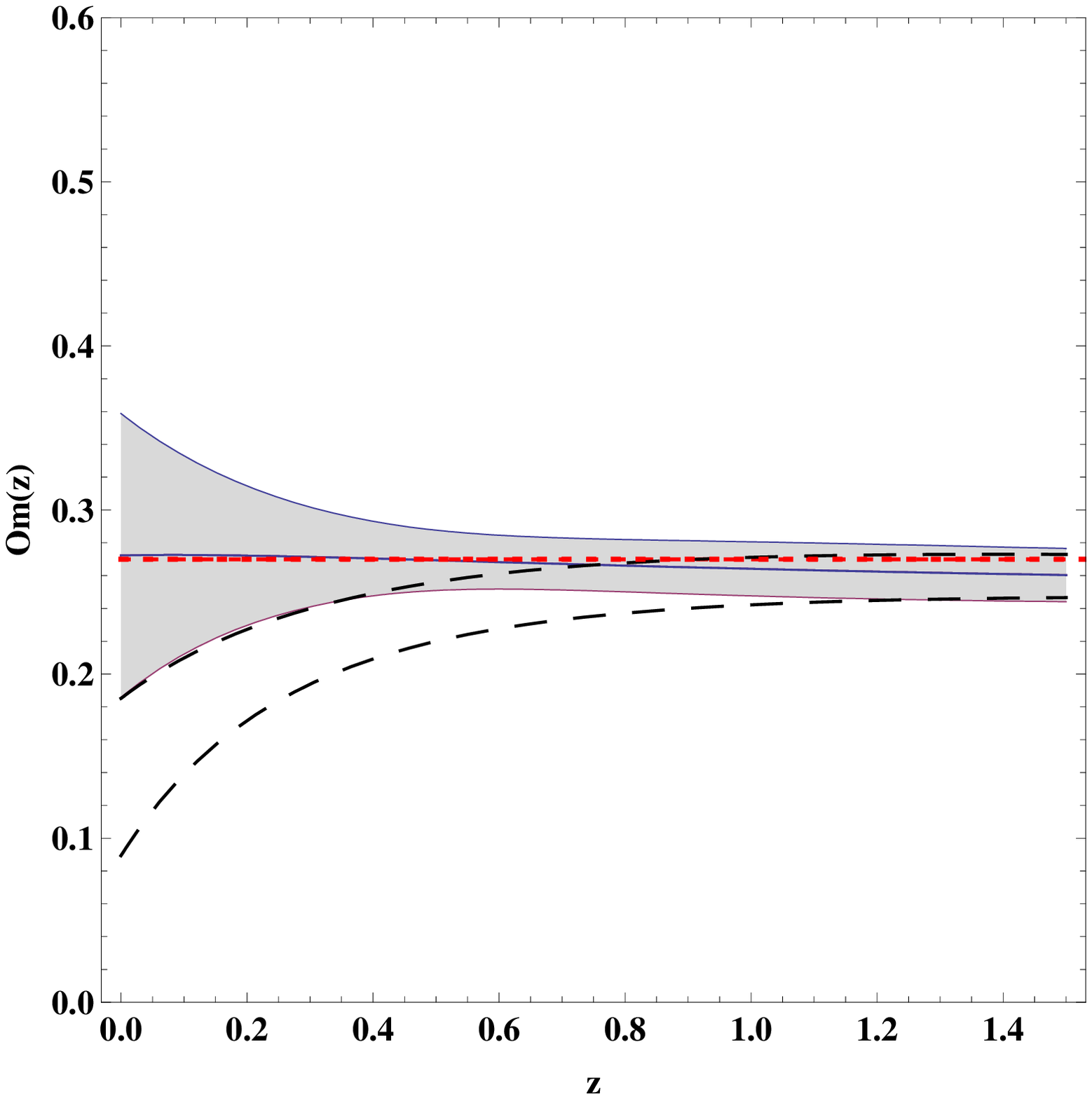}
 \caption{\label{Fig6}  The evolutionary behaviors of $q(z)$ and $Om(z)$ at the $68.3\%$
confidence level. A subsample (Union2T) of Union2 obtained with the
method in \cite{Nesseris2007} is considered.  The gray regions and
the regions between two long dashed lines  show the results without
and with the systematic errors in the SNIa, respectively.   The
upper and lower panels
 represent the results reconstructed  from Union2T+BAO and Union2T+BAO+CMB, respectively. }
 \end{figure}
 \end{center}

Let us now discuss  another method in selecting a subsample of the
SNIa data, which is based upon different considerations.  This
method was proposed by Nesseris and
Perivolaropoulos~\cite{Nesseris2007} to find the outliers
responsible for the tension in the SNIa data.  In this method, a
truncated version of the SNIa can be obtained by calculating the
relative deviation to the best fit $\Lambda$CDM prediction and
adopting a reasonable cut $|\mu_{obs}-\mu_{\Lambda
CDM}|/\sigma_{obs}$ beyond $1.9\; \sigma$. Using this method, we
find that  39 SNIa points distributed in the whole Union2 dataset
should be discarded. The names of these 39 SNIa are listed in
Tab.~(\ref{Tab2}). Thus, there remain 518 data points and we call
them ``Union2T". The results from Union2T are shown
Figs.~(\ref{Fig5}, \ref{Fig6}). We find, from Fig.~(\ref{Fig5}),
that the tension is also reduced significantly, and,  the
$\Lambda$CDM is consistent with Union2T(sys) and
Union2T(sys)+BAO+CMB at the $68\%$ confidence level.  If  the
systematic error is ignored, the observation favors an expansion
with an increasing acceleration at the present once the CMB is
added, although Union2T+BAO still support that the cosmic
acceleration is slowing down. This can been seen by looking at  the
grey regions in Fig.~(\ref{Fig6}). This result is similar with that
obtained from Union2, but is different from that from Union2S.
However, once the systematic errors are considered in the SNIa, both
Union2T(sys)+BAO and Union2T(sys)+BAO+CMB favor an expansion with an
increasing  acceleration at the present, which is different from
that from Union2 and Union2S. In Tab.~(\ref{Tab3}), we give the
$\chi^2/dof$ (dof: degree of freedom) value of different datasets,
from which, one can see that only in the case of Union2T is
$\chi^2/dof$ significantly improved. That is, according to the
$\chi^2/dof$ criterion,  the method proposed in ~\cite{Nesseris2007}
is preferred.

\begin{table}[!h]
\begin{tabular}{|c|}
\hline
The subset (39 SNIa) cut in Union2\\
\hline
1998dx, 1999bm, 2001v, 2002bf, 2002hd, 2002hu, 2002jy,
2003ch, 2003ic, 2006br, \\
2006cm, 2006cz, 2007ca, 10106, 2005ll, 2005lp, 2005fp, 2005gs,
2005gr, 2005hv, \\
~2005ig, 2005iu, 2005jj, 1997k, 2001jm, 1998ba, 03D4au, 04D3cp,
04D3oe,
03D4cx,~~\\
03D1co, d084, e140, f308, g050, g120, m138, 05Str, 2002fx\\
\hline
\end{tabular}
\tabcolsep 0pt \caption{\label{Tab2} The names of SNIa cut in the
Union2 with the method in Ref.~\cite{Nesseris2007}. } \vspace*{5pt}
\end{table}

\begin{table}[!h]
\begin{tabular}{|c|c||c|c|}
\hline
~~Dataset~~&~~$\chi^2/dof$ ~~&~~Dataset~~&~~$\chi^2/dof$ ~~\\
\hline
Union2+BAO &~~ 0.962~~&~~Union2(sys)+BAO~~&~~0.938~~\\
\hline
Union2+BAO+CMB~~&~~ 0.964~~&~~Union2(sys)+BAO+CMB~~&~~0.938~~\\
\hline
Union2S+BAO~~&~~ 0.956~~&~~Union2S(sys)+BAO~~&~~0.944~~\\
\hline
~Union2S+BAO+CMB~~&~~0.955~~&~~Union2S(sys)+BAO+CMB~~&~~0.942~~\\
\hline
~Union2T+BAO~&~~0.653~~&~~Union2T(sys)+BAO~&~~0.631~~\\
\hline
~Union2T+BAO+CMB~&~~0.653~~&~~Union2T(sys)+BAO+CMB~&~~0.630~~\\
\hline
\end{tabular}
\tabcolsep 0pt \caption{\label{Tab3} Summary of the $\chi^2/dof$
from different data sets.  } \vspace*{5pt}
\end{table}

\section{CONCLUSION}\label{sec4}
In this Letter, we have examined the cosmic expanding history from
the latest 558 Union2 SNIa together with BAO and CMB data. For the
SNIa, the data with and without the systematic error are analyzed
respectively. The popular CPL parametrization is considered. We
first find that, independent of whether or not the systematic error
is considered,  there exists a tension between low redshift data
(SNIa+BAO) and high redshift one (CMB), but for the case  with the
systematic error considered this tension is weaker than that from
the SNIa without. By reconstructing the curves of $q(z)$ and $Om(z)$
from Union2+BAO, we obtain that for both the SNIa with and without
the systematic error the cosmic acceleration has already peaked at
redshift $z\sim 0.3$ and is decreasing. However, when the CMB data
is added in our analysis, this result changes dramatically and the
observation favors a cosmic expansion with an increasing
acceleration, which further confirms the existence of the tension.

In order to reduce this tension, two different methods given in
Refs.~\cite{slowing, Nesseris2007} are considered. With the method
in~\cite{slowing}, we obtain a subsample of Union2 labeled as
Union2S, which is given by excluding the Gold data, the high $z$
Hubble Space Telescope data and the older SNIa data sets in Union2.
Thus 388 data points are left in Union2S.  Using Union2S,  we find
that the tension between SNIa+BAO and CMB is reduced markedly. For
the case  without the systematic error both Union2S+BAO and
Union2S+BAO+CMB favor a decreasing of the cosmic acceleration at $z
< 0.3$. However, once the systematic error is added, Union2S+BAO+CMB
support a present  increasing cosmic acceleration, although the
result from Union2S+BAO is similar with the case without the
systematic error. According to the method given in
\cite{Nesseris2007}, we cut 39 data points in Union2. Thus, a
subsample (Union2T) containing 518 data points is obtained. With
Union2T  the tension is also reduced noticeably. However, when the
systematic error is ignored, the results from reconstructing $q(z)$
and $Om(z)$ are similar to that given by Union2. Union2T+BAO favor
that the accelerating expansion of the Universe is slowing down,
while Union2T+BAO+CMB do not. If the systematic error is considered,
both Union2T+BAO and Union2T+BAO+CMB support a present increasing
cosmic acceleration.  Therefore, when  the systematic error in the
SNIa is ignored, Union2S and Union2T give totally different results
once the CMB is added, with one suggesting a slowing-down cosmic
acceleration, the other just the opposite, although both of them can
reduce the tension between low redshift data and high redshift one.
If  the systematic errors in the SNIa is considered, we find that
the similar results are obtained from Union2S and Union2T.  Both
Union2S+BAO+CMB and Union2T+BAO+CMB support an increasing of the
present cosmic acceleration. So, in order to have a clear-cut answer
on whether the cosmic acceleration is slowing down or not, we still
need to wait for  more consistent data and more reliable methods to
analyze them.

In Tab.~(\ref{Tab3}), the values of $\chi^2/dof$ from different
dataset are given. From which one can see that this value is
significantly improved when Union2T is used. Thus, the
$\chi^2_{Min}/dof$ criterion indicates that the method given in
\cite{Nesseris2007} is favored by observations.

 Finally, we must
point out that all our results obtained in the present Letter are
based on the CPL parametrization. If one uses a different
parameterization, as in Refs.~\cite{slowing}, the results might
change.

\section*{ACKNOWLEDGEMENTS}
This work was supported in part by the National Natural Science
Foundation of China under Grants Nos. 10775050, 10705055, 10935013
and 11075083,  Zhejiang Provincial Natural Science Foundation of
China under Grant No. Z6100077, the SRFDP under Grant No.
20070542002, the FANEDD under Grant No. 200922, the National Basic
Research Program of China under Grant No. 2010CB832803, the NCET
under Grant No. 09-0144, and the PCSIRT under Grant No. IRT0964.

\end{document}